\documentclass[10pt]{elsarticle}
\usepackage{epsfig}

\include{graphics}
\begin{document}

\title{Asymptotic Reissner-Nordstr\"{o}m black holes}

\author[SHZ,RIAAM]{S. H.~Hendi}

\address[SHZ]{Physics Department and Biruni Observatory, College of Sciences, Shiraz
University, Shiraz 71454, Iran}

\address[RIAAM]{Research Institute for Astrophysics and Astronomy of Maragha
(RIAAM), P.O. Box 55134-441, Maragha, Iran}

\begin{abstract}
We consider two types of Born-Infeld like nonlinear
electromagnetic fields and obtain their interesting black hole
solutions. The asymptotic behavior of these solutions is the same
as that of Reissner-Nordstr\"{o}m black hole. We investigate the
geometric properties of the solutions and find that depending on
the value of the nonlinearity parameter, the singularity covered
with various horizons.
\end{abstract}



\maketitle
\section{Introduction}
Standard Maxwell theory is confronted with the infinite
self-energy of charged point-like particles. The renormalization
procedure in quantum electrodynamics theory has been applied to
obtain a finite value, as deduced from the difference between two
infinite forms \cite{Ryder}. The Lagrangian of linear Maxwell
field, $L_{Max}=-\mathcal{F}=-F_{\mu \nu }F^{\mu \nu }$, has been
considered in various theories of gravity \cite{GravMax}. It is
well-known that the Reissner--Nordstr\"{o}m (RN) metric is the
spherically symmetric solution of Einstein-Maxwell gravity, which
corresponds to the charged static black hole of mass $M$ and
charge $Q$ \cite{RN}. Within gravitational field theories,
generalization of RN black holes to nonlinear electromagnetic
fields plays a crucial role since most of physical systems are
intrinsically nonlinear in the nature.

Among nonlinear electromagnetic fields the well-known Born-Infeld
(BI) theory is quite special as low energy limiting cases of
certain models of string theory \cite{BIString}. This classical
theory has been designed to regulate the self-energy of a
point-like charge \cite{BI}. Although coupling of the BI theory
with Einstein gravity has been studied in $1935$ \cite{Hoffmann},
in recent years, black hole solutions of BI theory are of interest
\cite{BIpaper}.

Other kinds of nonlinear electrodynamics, with various
motivations, in the
context of gravitational field have been investigated \cite%
{PMIpaper,Oliveira,Soleng}. In this paper, we consider
$(3+1)$-dimensional action of Einstein gravity with two kinds of
BI-like nonlinear electromagnetic fields. Generalization of
Maxwell field to nonlinear BI model provides powerful tools for
investigation of interesting black hole solutions. In addition, it
was found that all order loop corrections to gravity may be
regarded as a BI-like Lagrangian \cite{Fradkin85}. Considering the
BI Lagrangian
\begin{equation}
L_{BI}=4\beta ^{2}\left( 1-\sqrt{1+\frac{\mathcal{F}}{2\beta
^{2}}}\right) , \label{LBI}
\end{equation}
one can expand it for large values of nonlinearity parameter
($\beta \longrightarrow \infty $) to obtain
\begin{equation}
\left. L_{BI}\right\vert _{{Large\; }\beta
}=-\mathcal{F}+\frac{1}{\beta ^{2}}O(\mathcal{F}^{2}),
\label{LBIexpand}
\end{equation}
which confirms that $L_{BI}$\ reduces to the linear Maxwell
Lagrangian for $\beta \longrightarrow \infty $.

Moreover, from cosmological point of view, it was shown that the
early universe inflation may be explained with the nonlinear
electrodynamics \cite{BICosmology}. Also, one can connect the
nonlinear electrodynamics with the large scale correlated magnetic
fields observed nowadays \cite{{NLCosmology}}.

Motivated by all the above, we seek four dimensional solutions to
Einstein gravity with the BI-like Lagrangian. We consider a
recently proposed BI-like model \cite{HendiJHEP}, namely
Exponential form of nonlinear electromagnetic field (ENEF) and
Logarithmic form of nonlinear electromagnetic field (LNEF), whose
Lagrangians are
\begin{equation}
L(\mathcal{F})=\left\{
\begin{array}{ll}
\beta ^{2}\left( \exp (-\frac{\mathcal{F}}{\beta ^{2}})-1\right) ,
& \;~{ENEF} \\
-8\beta ^{2}\ln \left( 1+\frac{\mathcal{F}}{8\beta ^{2}}\right) ,
& \;~{LNEF}
\end{array}
\right. .  \label{Lnon}
\end{equation}
in which, for large values of $\beta $, $\mathcal{F}^{2}$
correction to Maxwell Lagrangian is included. In other words, the
two forms of the mentioned Lagrangians have the same expansion as
presented in Eq. (\ref{LBIexpand}).

The plan of the paper is as follows. We give a brief review of the
simplest coupling of the nonlinear gauge field system (\ref{Lnon})
to Einstein gravity. Then, we obtain the solutions of the four
dimensional spherical symmetric spacetime in the presence of two
classes of nonlinear electromagnetic fields, investigate their
properties and compare them with BI solution. The paper ends with
some conclusions.

\section{$(3+1)$-dimensional black holes with nonlinear electromagnetic field}

Let us begin with the $(3+1)$-dimensional nonlinear
electromagnetic field in general relativity, with the action
\begin{equation}
I_{G}=-\frac{1}{16\pi }\int_{\mathcal{M}}d^{4}x\sqrt{-g}\left[
R-2\Lambda +L(\mathcal{F})\right] -\frac{1}{8\pi }\int_{\partial
\mathcal{M}}d^{3}x\sqrt{-\gamma }K,  \label{Act}
\end{equation}
where ${R}$ is the Ricci scalar, $\Lambda $ refers to the negative
cosmological constant and $L(\mathcal{F})$ is defined in Eq.
(\ref{Lnon}). In addition, the second integral in Eq. (\ref{Act})
is the boundary term that does not affect the equations of motion,
namely the Gibbons-Hawking boundary term \cite{GibHaw}. The
factors $K$ and $\gamma $ are, respectively, the traces of the
extrinsic curvature and the induced metric of boundary ${\partial
\mathcal{M}}$ of the manifold ${\mathcal{M}}$. The Einstein and
electromagnetic field equations derived from the above action are
\begin{equation}
R_{\mu \nu }-\frac{1}{2}g_{\mu \nu }\left( R-2\Lambda \right)
=\alpha \left( \frac{1}{2}g_{\mu \nu }L(\mathcal{F})-2F_{\mu
\lambda }F_{\nu }^{\;\lambda }L_{\mathcal{F}}\right) ,
\label{FE1}
\end{equation}
\begin{equation}
\partial _{\mu }\left( \sqrt{-g}L_{\mathcal{F}}F^{\mu \nu }\right) =0,
\label{FE2}
\end{equation}
where $L_{\mathcal{F}}=\frac{dL(\mathcal{F})}{d\mathcal{F}}$. Our
main aim here is to obtain charged static black hole solutions of
the field equations
(\ref{FE1}) and (\ref{FE2}) and investigate their properties. We assume a $%
(3+1)$-dimensional static spacetime is described by the
spherically symmetric line element
\begin{equation}
ds^{2}=-f(r)dt^{2}+\frac{dr^{2}}{f(r)}+r^{2}\left( d\theta
^{2}+\sin ^{2}\theta d\phi ^{2}\right) .  \label{Metric}
\end{equation}

The electromagnetic tensor $F_{\mu \nu }$ compatible with the
metric (\ref{Metric}) can involve only a radial electric field
$F_{tr}=-F_{rt}$. Therefore, we should use the gauge potential
ansatz $A_{\mu }=h(r)\delta _{\mu }^{t}$ in the nonlinear
electromagnetic field equation (\ref{FE2}). Considering the
mentioned ansatz, one can show that Eq. (\ref{FE2}) reduces to
\begin{equation}
\begin{array}{rr}
r\left[ 1+\left( \frac{2h^{\prime }(r)}{\beta }\right) ^{2}\right]
h^{\prime
\prime }(r)+2h^{\prime }(r)=0, & \;~{ENEF}\vspace{0.1cm} \\
r\left[ 1+\left( \frac{h^{\prime }(r)}{2\beta }\right) ^{2}\right]
h^{\prime \prime }(r)+2h^{\prime }(r)\left[ 1-\left(
\frac{h^{\prime }(r)}{2\beta } \right) ^{2}\right] =0, & \;~{LNEF}
\end{array},  \label{heq}
\end{equation}
with the following solutions
\begin{equation}
h(r)=\left\{
\begin{array}{ll}
-\frac{1}{10}\left( 4Q^{2}\beta ^{2}L_{W}e^{-L_{W}}\right)
^{1/4}\left[ 5+L_{W} F {\left( \left[ 1\right] ,\left[
\frac{9}{4}\right] ,\frac{{Lw}}{4}\right) }\right], & \;~{ENEF}\vspace{0.1cm} \\
\frac{2Q}{3r}\left[ \frac{1}{\left( 1+\Gamma \right) }-2 F {\left(
\left[ \frac{1}{4},\frac{1}{2}\right] ,\left[ \frac{5}{4}\right]
,1-\Gamma ^{2}\right) }\right] , & \;~{LNEF}
\end{array}
\right. ,  \label{h(r)}
\end{equation}
where prime and double primes denote the first and second
derivative with respect to $r$, respectively, $\Gamma
=\sqrt{1+\frac{Q^{2}}{r^{4}\beta ^{2}}} $ and $Q$ is an
integration constant which is the electric charge of the black
holes. In addition, $L_{W}=LambertW\left( \frac{4Q^{2}}{\beta
^{2}r^{4}}\right) $ which satisfies $LambertW(x)\exp \left[
LambertW(x)\right] =x$ and $F ([a],[b],z)$ is hypergeometric
function (for more details, see \cite{Lambert}).

It is notable that the same procedure for the linear Maxwell and
nonlinear BI theory leads to
\begin{equation}
h(r)=\left\{
\begin{array}{ll}
-\frac{Q}{r}, & \;~{Maxwell}\vspace{0.1cm} \\
-\frac{Q}{r}F {\left( \left[ \frac{1}{4},\frac{1}{2}\right]
,\left[ \frac{5}{4}\right] ,1-\Gamma ^{2}\right) }, &
\;~{BINEF}\vspace{0.1cm}
\end{array}
\right.  \label{h(r)2}
\end{equation}
which are completely different with Eq. (\ref{h(r)}). In order to
find the asymptotic behavior of the solutions, we should expand
them for large distances ($r>>1$). The radial function $h(r)$ is
expanded as
\begin{equation}
\left. h(r)\right\vert _{{Large \;}r}=-\frac{Q}{r}+\frac{\chi
Q^{3}}{20 \beta ^{2} r^{5}}+O\left( \frac{1}{r^{9}}\right) ,
\label{hexpand}
\end{equation}
where $\chi =2$, $8$ and $1$ for BINEF, ENEF and LNEF,
respectively. So, we find that for large values of $r$, the
dominant term of Eq. (\ref{hexpand}) is the gauge potential of RN
black hole. Now, we consider the nonvanishing components of
$F_{\mu \nu }$. One can show that
\begin{equation}
F_{tr}=-F_{rt}=\frac{Q}{r^{2}}\times \left\{
\begin{array}{ll}
e^{-\frac{L_{W}}{2}}, & \;~{ENEF} \\
\frac{2}{\Gamma +1}, & \;~{LNEF}%
\end{array}
\right. ,  \label{Ftr}
\end{equation}
and it has been shown for BI theory \cite{BIpaper}
\[
F_{tr}=-F_{rt}=\frac{Q}{\Gamma r^{2}}.
\]
Differentiating from Eq. (\ref{hexpand}) or expanding $F_{tr}$ for
large distances, one can obtain
\begin{equation}
F_{tr}=\frac{Q}{r^{2}}-\frac{\chi Q^{3}}{4\beta ^{2}r^{6}}+O\left(
\frac{1}{r^{10}}\right) ,  \label{Ftrexpand}
\end{equation}
which contain additional radial electric fields apart from the
usual Coulomb one.

It is easy to find that all the mentioned electric fields vanish
at large values of $r$, as they should be. Furthermore, one may
expect to obtain a finite value for the nonlinear electric fields
at $r=0$. It is interesting to mention that, despite ENEF, the
electric field of LNEF is finite at the origin. One can find that
$F_{tr}^{ENEF}$ has a finite value \emph{near} the origin, which
is the same as the electric fields of BINEF and LNEF, but
$F_{tr}^{ENEF}$ diverges at $r=0$, as it occurs for Maxwell field.

Now, we should find a suitable function $f(r)$ to satisfy all
components of Eq. (\ref{FE1}). Considering Eq. (\ref{h(r)}), one
may show that the $rr$ and $tt$ components of Eq. (\ref{FE1}) can
be simplified as
\begin{equation}
Eq_{tt}=Eq_{rr}=f^{\prime }(r)+\frac{f(r)+\Lambda
r^{2}-1}{r}-r\beta ^{2}g(r)=0,  \label{f(r)eq}
\end{equation}
where
\begin{equation}
g(r)=\left\{
\begin{array}{ll}
1-\frac{Q}{\beta r^{2}\sqrt{L_{W}}}\left( 1-L_{W}\right) , &
\;~{ENEF}\vspace{0.2cm} \\
4\left[ 1-\Gamma -\ln \left( \frac{2}{\Gamma +1}\right) \right] ,
& \;~{LNEF}
\end{array}
\right. .  \label{g(r)}
\end{equation}
Other nonzero components of Eq. (\ref{FE1}) can be written as
\begin{equation}
Eq_{\theta \theta }=Eq_{\phi \phi }=\left(
\frac{d}{dr}-\frac{1}{r}\right) Eq_{tt}=0,  \label{f(r)eq2}
\end{equation}
and therefore it is sufficient to solve $Eq_{tt}=0$. After some
cumbersome calculations, the solutions of Eq. (\ref{f(r)eq}) can
be written as
\begin{equation}
f(r)=1-\frac{2M}{r}-\frac{\Lambda r^{2}}{3}+\left\{
\begin{array}{ll}
\frac{Q}{3}\left( \frac{4Q^{2}\beta
^{6}e^{-L_{W}}}{r^{4}L_{W}^{5}}\right) ^{1/8}\left[
1+L_{W}+\frac{4}{5}L_{W}^{2} F {\left( \left[ 1\right] , \left[
\frac{9}{4}\right] ,\frac{L_{W}}{4}\right) }\right] , & \;~{ENEF}
\vspace{0.2cm} \\
\frac{8Q^{2} F {\left( \left[ \frac{1}{2},\frac{1}{4}\right]
,\left[ \frac{5}{4}\right] ,1-\Gamma ^{2}\right)
}}{3r^{2}}-\frac{4\beta ^{2}r^{2}\left( \Gamma -\ln \left[
\frac{\left( \Gamma ^{2}-1\right) }{2e^{-7/3}}\right] \right)
}{3}-\frac{4\beta ^{2}\int r^{2}\ln \left( \Gamma -1\right)
dr}{r}, & \;~{LNEF}
\end{array}
\right. ,  \label{F(r)}
\end{equation}
where $M$ is the integration constant which is the total mass of
spacetime (the integral term of Eq. (\ref{F(r)}) will be solved in
the appendix). In order to compare our solutions with the BI and
RN black holes, we should present the BI and RN solutions.
Considering Eq. (\ref{h(r)2}), it has been shown that the field
equation (\ref{FE1}) reduces to the Eqs. (\ref{f(r)eq}) and
(\ref{f(r)eq2}) with
\begin{equation}
g(r)=\left\{
\begin{array}{ll}
\frac{Q^{2}}{r^{3}}, & \;~{RN}\vspace{0.2cm} \\
2\left( 1-\Gamma \right) , & \;~{BINEF}\vspace{0.2cm}
\end{array}
\right. ,  \label{g(r)2}
\end{equation}
and the following solutions
\begin{equation}
f(r)=1-\frac{2M}{r}-\frac{\Lambda r^{2}}{3}+\left\{
\begin{array}{ll}
\frac{Q^{2}}{r^{2}}, & \;~{RN}\vspace{0.2cm} \\
\frac{4Q^{2}}{3r^{2}} F {\left( \left[
\frac{1}{4},\frac{1}{2}\right] , \left[ \frac{5}{4}\right]
,1-\Gamma ^{2}\right) }, & \;~{BINEF}\vspace{
0.2cm}%
\end{array}
\right. .  \label{F(r)2}
\end{equation}
We should note that although Eq. (\ref{F(r)}) is different with RN
and BI solutions, as previously mentioned, they display the same
asymptotic behavior. The metric functions of nonlinear
electromagnetic fields are expanded as
\begin{equation}
f(r)=1-\frac{2M}{r}-\frac{\Lambda
r^{2}}{3}+\frac{Q^{2}}{r^{2}}-\frac{\chi Q^{4}}{40r^{6}\beta
^{2}}+O\left( \frac{1}{r^{10}}\right) . \label{f(r)expand}
\end{equation}
Considering the first correction term in Eq. (\ref{f(r)expand}),
we should stress that the this equation contains both the usual RN
terms as well as an analytic function of $O(\frac{1}{r^{6}})$, in
which one can ignore it to obtain RN solution for large values of
$r$.

In what follows, we are going to investigate the geometric nature
of the obtained solutions. At first, we look for the existence of
curvature singularities and their horizons. Given the metric in
Eq. (\ref{Metric}), we can compute the Ricci and the Kretschmann
scalars
\begin{eqnarray}
R &=&-f^{\prime \prime }(r)-\frac{4f^{\prime
}(r)}{r}-\frac{2f(r)}{r^{2}},
\label{R} \\
R_{\mu \nu \rho \sigma }R^{\mu \nu \rho \sigma } &=&f^{\prime
\prime 2}(r)+\left( \frac{2f^{\prime }(r)}{r}\right) ^{2}+\left(
\frac{2f(r)}{r^{2}}\right) ^{2}.  \label{RR}
\end{eqnarray}
where $f(r)$ is the metric function. Also one can show that other
curvature invariants (such as Ricci square) are functions of
$f^{\prime \prime }$, $f^{\prime }/r$ and $f/r^{2}$, and therefore
it is sufficient to study the Ricci and the Kretschmann scalars
for the investigation of the spacetime curvature. After a number
of manipulations, we find these scalars diverge in the vicinity of
the origin
\begin{eqnarray}
\lim_{r\longrightarrow 0^{+}}R &=&\infty ,  \label{Rorigin} \\
\lim_{r\longrightarrow 0^{+}}R_{\mu \nu \rho \sigma }R^{\mu \nu
\rho \sigma } &=&\infty ,  \label{RRorigin}
\end{eqnarray}
which confirms that the spacetime given by Eqs. (\ref{Metric}) and
(\ref{F(r)}) has an essential singularity at $r=0$.

\begin{figure}[h]
\centering
\includegraphics[width=7cm]{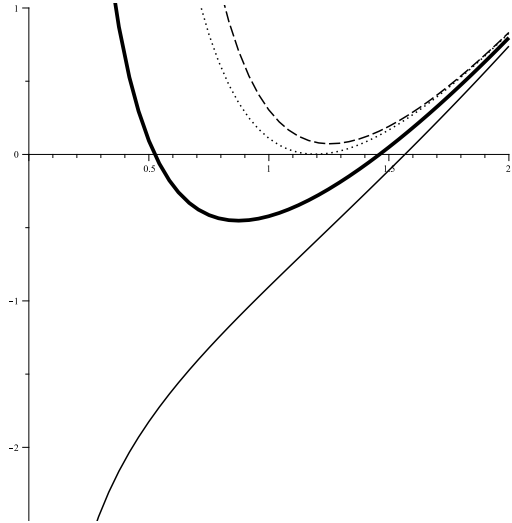}
\caption{$f(r)$ (ENEF branch) versus $r$ for $\Lambda=-1$, $q=2$,
$M=2.5$, and $\protect\beta=0.5<\protect\beta_{c}$ (solid line:
black hole with a non-extreme horizon),
$\protect\beta_{c}<\protect\beta=1<\protect\beta_{ext}$ (bold
line: black hole with two horizons),
$\protect\beta=\protect\beta_{ext}=3$ (dotted line: black hole
with an extreme horizon) and
$\protect\beta=10>\protect\beta_{ext}$ (dashed line: naked
singularity) } \label{Metricexp}
\end{figure}
\begin{figure}[h]
\centering
\includegraphics[width=8cm]{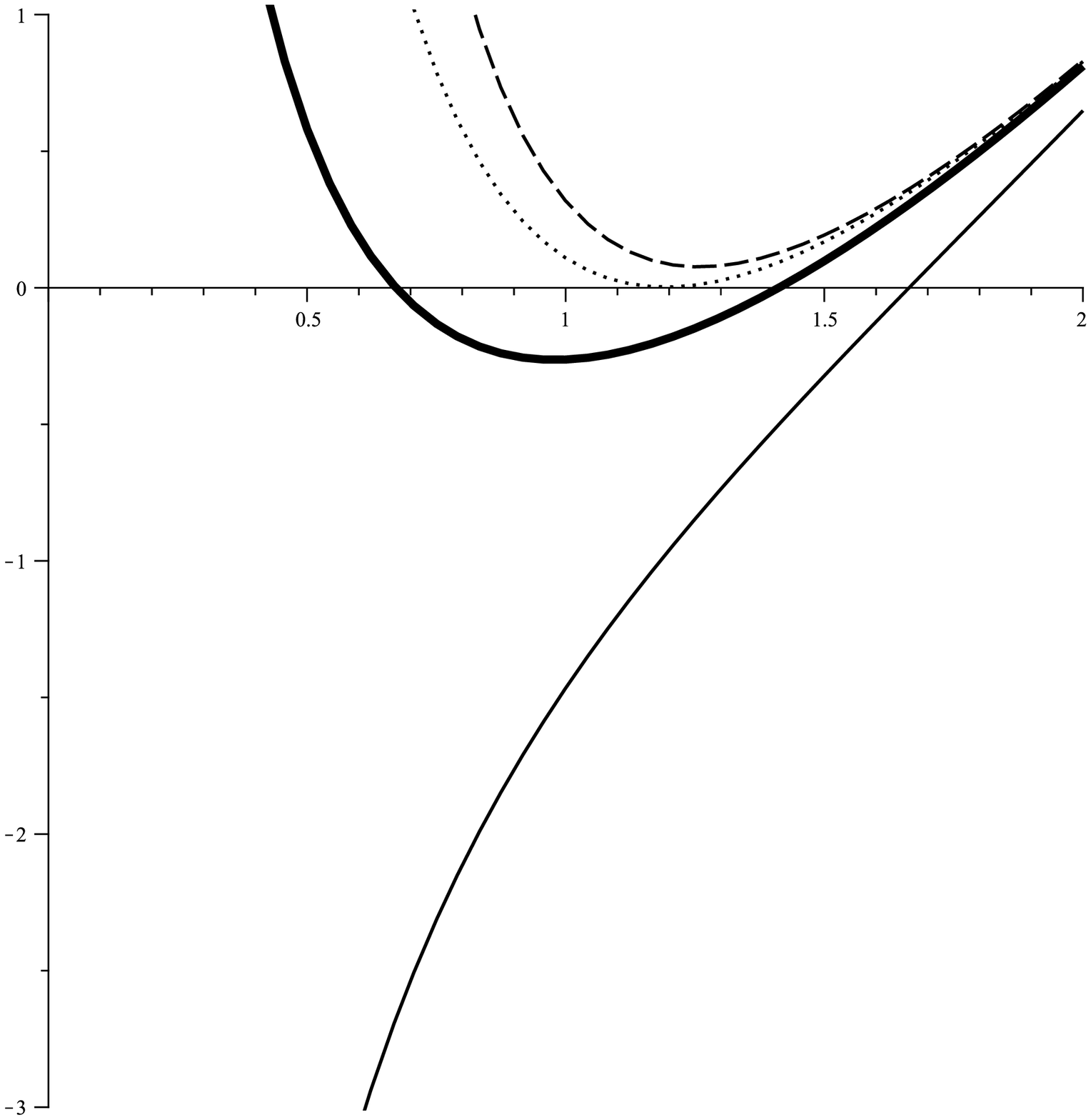}
\caption{$f(r)$ (LNEF branch) versus $r$ for $\Lambda=-1$, $q=2$,
$M=2.5$, and $\protect\beta=0.1<\protect\beta_{c}$ (solid line:
black hole with a non-extreme horizon),
$\protect\beta_{c}<\protect\beta=0.5<\protect\beta_{ext}$ (bold
line: black hole with two horizons),
$\protect\beta=\protect\beta_{ext}=1.1$ (dotted line: black hole
with an extreme horizon) and $\protect\beta=5>\protect\beta_{ext}$
(dashed line: naked singularity) } \label{Metriclog}
\end{figure}

In order to investigate the asymptotic behavior of the solution,
we should simplify Eqs. (\ref{R}) and (\ref{RR}) for large
distances. Expanding these equations and keep the first order
nonlinear correction, we can obtain
\begin{eqnarray}
\lim_{r\longrightarrow \infty }R &=&4\Lambda +\frac{\chi
Q^{4}}{2\beta
^{2}r^{8}}+O\left( \frac{1}{r^{12}}\right) ,  \label{Rinf} \\
\lim_{r\longrightarrow \infty }R_{\mu \nu \rho \sigma }R^{\mu \nu
\rho \sigma } &=&\frac{8}{3}\Lambda
^{2}+\frac{48M^{2}}{r^{6}}-\frac{96MQ^{2}}{r^{7}}+\frac{56Q^{4}}{r^{8}}-\frac{2\chi
Q^{4}}{\beta ^{2}l^{2}r^{8}} +O\left( \frac{1}{r^{11}}\right) ,
\label{RRinf}
\end{eqnarray}
where the last term is the leading nonlinear correction to the RN
black hole solutions. Equations (\ref{Rinf}) and (\ref{RRinf})
show that the asymptotic behavior of the obtained solutions is adS
for $\Lambda <0$.

Now, we investigate the effects of nonlinear electromagnetic
fields and the nonlinearity parameter, $\beta$. Figs.
\ref{Metricexp} and \ref{Metriclog} show that the nonlinearity
parameter has effect on the value of the inner and outer horizons
in addition to the minimum value of the metric function (when
$f(r)$ has a minimum). Furthermore, considering Eq. (\ref{F(r)2}),
we find that due to the fact that the metric function of RN black
hole is positive near the origin as well as large values of $r$,
depending on the value of the metric parameters, one can obtain a
black hole with two inner and outer horizons, an extreme black
hole with zero temperature and a naked singularity. It is so
interesting to mention that for the nonlinear solutions, a new
situation may appear. In other words, one can show that near the
origin ($r\longrightarrow 0$), the metric function, Eq.
(\ref{F(r)}), may be positive, zero or negative for $\beta >\beta
_{c}$, $\beta =\beta _{c}$ or $\beta <\beta _{c}$, respectively
(see Figs. \ref{Metricexp} and \ref{Metriclog} for more details).
We should note that considering $\lim_{r\longrightarrow
0^{+}}f(r)=0$, one may obtain $\beta _{c}$ as a function of
$\Lambda$, $Q$ and $M$, numerically. The new situation appears for
$\beta <\beta _{c}$, in which the black holes have one non-extreme
horizon with positive temperature as it happens for (adS)
Schwarzschild solutions (uncharged solutions). In addition, Figs.
\ref{Metricexp} and \ref{Metriclog} show that for $\beta >\beta
_{c}$, we may obtain an extreme value for the nonlinearity
parameter ($\beta _{ext}$) to achieve a black hole with two
horizons, an extreme black hole and a naked singularity for $\beta
_{c}<\beta <\beta _{ext}$, $\beta =\beta _{ext}$ and $\beta
>\beta _{ext}$, respectively.

\section{Conclusions} \label{sectconcl}

In this paper, we have considered two classes of the BI-like
nonlinear electromagnetic fields in the Einsteinian gravity.
Expansion of both the mentioned nonlinear Lagrangians, for large
values of nonlinearity parameter, is the same as that of BI theory
and so we called them BI-like fields.

Regarding four dimensional spherically symmetric spacetime, we
have obtained the compatible electromagnetic fields and found
that, in contrast with Maxwell field, the mentioned nonlinear
electromagnetic fields have finite values in the vicinity of the
origin. In addition, we have found that $F_{tr}^{ENEF}$ diverges
at $r=0$, but its divergency is much slower than the Maxwell one.
In other words, the behavior of the $F_{tr}^{LNEF}$ is very close
to BI field, but in the neighborhood of the origin,
$F_{tr}^{ENEF}$ has a special behavior between $F_{tr}^{BINEF}$
and $F_{tr}^{RN}$. Expansion of the nonlinear electromagnetic
fields for large $r$ showed that the asymptotic behavior of all of
them is exactly the same as linear Maxwell field.

Then, we have solved the gravitational field equations and
obtained asymptotic AdS solutions with a curvature singularity at
$r=0$. We have found that depending on the values of the
nonlinearity parameter, $\beta $, the singularity is covered with
various horizons. To state the matter
differently, we should note that one can obtain two nonlinearity values, $%
\beta _{c}$ (critical value) and $\beta _{ext}$ (extreme value),
in which the singularity covered with a non-extreme horizon for
$\beta <\beta _{c}$,
two horizons when $\beta _{c}<\beta <\beta _{ext}$, an extreme horizon for $%
\beta =\beta _{ext}$, and there is a naked singularity otherwise.
We have plotted some figures for more clarifications. Finally, we
should remark that the ($3+1$)-dimensional nonlinear charged black
holes described here have similar asymptotic properties to RN
black holes which are well studied.

It would be interesting to carry out a study of the thermodynamic
and dynamic properties of the mentioned solutions. Also, it is
worthwhile to generalize these four dimensional solutions to the
rotating and higher dimensional cases. We leave these problems for
future works.

\section*{Acknowledgements}
We also wish to thank Shiraz University Research Council. This
work has been supported financially by Research Institute for
Astronomy \& Astrophysics of Maragha (RIAAM), Iran.

\section*{Appendix}

In order to complete our discussion, we may solve the following
integral
\begin{eqnarray*}
\int r^{2}\ln \left( \Gamma -1\right) dr &=&-\frac{q^{3/2}(\Gamma
-1)^{1/4}}{2^{3/4}\beta ^{3/2}}\left[ \frac{14}{3} F {\left(
\left[ \frac{1}{4},\frac{1}{4},\frac{11}{4}\right] ,\left[
\frac{5}{4},\frac{5}{4}\right] ,\frac{1-\Gamma }{2}\right) }-\right. \\
&&\left. \frac{14}{25}(\Gamma -1) F {\left( \left[
\frac{5}{4},\frac{5}{4},\frac{11}{4}\right] ,\left[
\frac{9}{4},\frac{9}{4}\right] ,\frac{1-\Gamma }{2}\right) -}\right. \\
&&\left. \frac{\left[ 4+3\ln (\Gamma -1)\right] }{9(\Gamma -1)} F
{\left( \left[ \frac{-3}{4},\frac{7}{4}\right] ,\left[
\frac{1}{4}\right] ,\frac{1-\Gamma }{2}\right) }+\right. \\
&&\left. \left[ -4+\ln (\Gamma -1)\right]  F {\left( \left[
\frac{1}{4},\frac{7}{4}\right] ,\left[ \frac{5}{4}\right]
,\frac{1-\Gamma }{2}\right)} \right].
\end{eqnarray*}

\end{document}